\documentclass[12pt,preprint]{aastex}

\newcommand{\EQ}{\begin{equation}}
\newcommand{\EN}{\end{equation}}
\newcommand{\EQA}{\begin{eqnarray}}
\newcommand{\ENA}{\end{eqnarray}}
\usepackage{amsmath}
\newcommand{\A}{\mathcal{A}}

\newcommand{\ie}{{\em i.e. }}

\begin{document}

\title{Effect of Rossby and Alfv\'en waves on the dynamics of the tachocline}

\author{Nicolas Leprovost \and Eun-jin Kim}

\affil{Department of Applied Mathematics, University of Sheffield, Sheffield S3 7RH, UK}

\begin{abstract}
To understand magnetic diffusion, momentum transport, and mixing in the interior of the sun, we consider an idealized model of the tachocline, namely magnetohydrodynamics (MHD) turbulence on a $\beta$ plane subject to a large scale shear (provided by the latitudinal differential rotation). This model enables us to self-consistently derive the influence of shear, Rossby and Alfv\'en waves on the transport properties of turbulence. In the strong magnetic field regime, we find that the turbulent viscosity and diffusivity are reduced by magnetic fields only, similarly to the two-dimensional MHD case (without Rossby waves). In the weak magnetic field regime, we find a crossover scale ($L_R$) from a Alfv\'en dominated regime (on small scales) to a Rossby dominated regime (on large scales). For parameter values typical of the tachocline, $L_R$ is larger that the solar radius so that Rossby waves are unlikely to play an important role in the transport of magnetic field and angular momentum. This is mainly due to the enhancement of magnetic back-reaction by shearing which efficiently generates small scales, thus strong currents.
\end{abstract}

\keywords{Turbulence -- Sun: interior -- Sun: rotation}

\section{Introduction}
Data from global helioseismology \citep{Thompson03} have shed some light on the internal rotation of the sun. Throughout the convective envelope, the rotation rate decreases monotonically toward the poles. Near the base of the convection zone, there is a sharp transition between differential rotation in the convective envelope and nearly uniform rotation in the radiative interior. This transition region has become known as the solar tachocline \citep{Spiegel72}, and its thickness has been constrained to be only a few percent of the solar radius by helioseismic data \citep{Charbonneau99}. The rotation rate of the radiative interior is intermediate between the equatorial and polar regions of the convection zone. Thus, the radial angular velocity gradient across the tachocline is positive at low latitudes and negative at high latitudes. The tachocline is most probably located below the convection zone \citep{Charbonneau99}, but part of it might be included in the overshooting region where turbulent plumes coming from the convection zone can induce turbulent motions. This overshooting layer contains a prominent latitudinal differential rotation. It is thus of primary importance to understand the physics pertinent to such a turbulent layer in presence of strong shear. 
In particular, as the tachocline links two regions of very different transport properties, theoretical predictions of transport coefficients are essential to understand its long term dynamics (e.g. its thickness). For instance, different mechanisms have been invoked to stabilize this transition layer against the radiative broadening during the solar evolution. First, since the radial mixing is ineffective due to the stable stratification in the tachocline, \citet{Spiegel92} suggested that the efficient turbulent transport of angular momentum (eddy-viscosity) in the horizontal direction could be responsible for the confinement of the tachocline. Other models rely on the existence of a magnetic field fully contained in the radiative zone \citep{Rudiger96,Gough98,MacGregor99}. Specifically, \citet{Rudiger96} have shown that this configuration could explain the dynamics of the tachocline if the effective viscosity is larger by at least 4 orders of magnitude than its molecular value. 
\citet{Gough98} and \citet{MacGregor99} have estimated the thickness of the tachocline by balancing the shearing of the poloidal field with the resistive dissipation of toroidal magnetic field. However, in this estimate, the turbulent diffusivity and viscosity might be more relevant than the molecular values. The tachocline is also important for the generation of magnetic field inside the sun (by means of a dynamo process), whose operation crucially depends on the values of effective (eddy) viscosity and magnetic diffusivity \citep[e.g.][]{Parker93}.

The purpose of this paper is to provide a consistent theory of momentum transport and magnetic field diffusion in the tachocline. In view of the tendency of two-dimensional (2D) turbulence in a stably stratified layer, we focus on the dynamics in the (local) horizontal plane orthogonal to density gradient, by incorporating the effects of rotation and toroidal magnetic fields. Specifically, the latitudinal differential rotation is represented by a large-scale shear flow while the latitudinal variation in the Coriolis force is captured by using a local $\beta$ plane. This model permits us to study the influence on turbulent transport of shear flow, Rossby and Alfv\'en waves. The shear flow efficiently generates small scales by shearing, thereby enhancing the back-reaction of fluctuating magnetic fields. We show that in the limit of strong magnetic field, the turbulent viscosity and magnetic diffusivity are reduced by magnetic fields only, with the same results as in the 2D magnetohydrodynamics (MHD) case \citep{Kim01}. In the weak magnetic field regime, we find a crossover from a Alfv\'en dominated regime (on small scales) to a Rossby dominated regime (on large scales). Using parameter values typical of the tachocline, we identify these two regimes and show that Rossby waves are unlikely to play an important role in the tachoclinic turbulent transport. This work complements our previous study of 3D hydrodynamical (HD) turbulence in the tachocline \citep{Kim05,2Shears}, within which the effects of shear flows (due to radial and latitudinal differential rotations) and the average rotation were investigated.  

The remainder of the paper is organised as follows: we provide our model of the tachocline and the governing equations in section \ref{Model}, the results for the hydrodynamical (HD) and  magnetohydrodynamical (MHD) cases in section \ref{Hydrodynamics} and \ref{MHD} respectively. In section \ref{Conclusion}, we discuss the implication of our findings for the tachocline.

\section{Model}
\label{Model}
To model the tachocline, we consider the motion in a plane ($y$,$z$) independent of the third coordinate $x$. In the tachocline case, the $x$, $y$ and $z$ directions stand respectively for the radial, azimuthal and latitudinal directions. For simplicity, we assume that the Coriolis force varies linearly with latitude: ${\bf \Omega} = (\Omega_0 + \beta z) {\bf \hat{x}}$. This model has originally been introduced by \citet{Rossby39} to study a homogeneous sheet of fluid on a sphere, the effect of the sphericity being only in the variation of the rotation rate with latitude. The term containing $\beta$ is responsible for the appearance of the so-called Rossby waves. We also assume that, on large-scale, there are a linear shear flow (due to latitudinal differential rotation), ${\bf U_0} = - (\A z) {\bf \hat{y}}$, and a uniform toroidal magnetic field, ${\bf B_0} = B_0 {\bf \hat{y}}$, where $\A > 0$ and $B_0 > 0$ without loss of generality. This magnetic field supports Alfv\'en waves in the system whereas a shear flow will favor the creation of small scale in the direction of the shear and thus enhance the dissipation \citep{Kim01,Kim03,Kim06}. The generation of small scales by a shear flow enhances the effect of Lorentz force via the formation of strong current. Consequently Alfv\'en waves tend to become dynamically more important than Rossby waves, as shall be shown shortly. 

Owing to the geometry of the problem, it is convenient to work with the vorticity of the fluctuations $\omega = ({\bf \nabla \times v}) \cdot {\bf \hat{x}} $ and their magnetic potential defined by ${\bf b} = {\bf \nabla} \times (a {\bf \hat{x}})$. We study these fluctuations in the framework of the quasi-linear approximation \citep{Moffatt78} where the interaction between the small scale fields can be neglected:
\EQA
\label{System1}
(\partial_t + U_0 \partial_y) \omega &=& - \beta v_z - (B_0 \partial_y) \nabla^2 a + \nu \nabla^2 \omega + f_w  \; , \\ \nonumber
 (\partial_t + U_0 \partial_y) a &=& - v_z B_0 + \eta \nabla^2 a  \; .
\ENA
Here, we assume that the turbulence is driven mainly hydrodynamically (no force in the equation for the magnetic field). 
For simplicity, we assume an unit magnetic Prandtl number ($\nu = \eta$) and introduce a time dependent Fourier transform to capture the shearing effect:
\begin{equation}
\nonumber
F({\bf x},t) = \frac{1}{(2\pi)^2} \int d^2 {\bf k} \; e^{i[k_y y + { k_z(t)} z]}
\tilde{F}({\bf k},t) \; ,
\end{equation}
where $F$ stands for $\omega$, $a$ or $f_w$ and the azimuthal wave number evolves as: $\partial_t k_z = k_y \A$. By using the transformation: $\hat{F} = \tilde{F} \exp[\nu(k_y^2 t + k_z^3/3k_y \A)]$ to absorb the diffusion term and by transforming the time variable from $t$ to $\tau = k_z(t) / k_y = k_z(t_0) / k_y + \A (t-t_0)$, Eq. (\ref{System1}) can be rewritten:
\EQA
\label{System2}
\partial_\tau \hat{\omega} &=& i  \frac{\alpha \hat{\omega}}{(1+\tau^2)}  + i k_y^2 \gamma (1+\tau^2) \hat{a} + \frac{\hat{f_w}}{\A}  \; , \\ \nonumber
\partial_\tau \hat{a} &=& i \frac{\gamma \hat{\omega}}{k_y^2(1+\tau^2)}   \; .
\ENA
Here, $\alpha = \beta / \A k_y$ is the ratio between the rate of change in the rotation vector and the shearing rate and $\gamma = B_0 k_y / \A$ is the ratio between the magnetic field and the shear. Note that the velocity and magnetic fields can be obtained from the vorticity and the magnetic potential by the identities: $\hat{v}_y = i \frac{k_z \hat{\omega}}{k^2}$, $\hat{v}_z = - i \frac{k_y \hat{\omega}}{k^2}$, $\hat{b}_y = i k_z \hat{a}$  and $\hat{b}_z = - i k_y \hat{a}$. Equation (\ref{System2}) can be solved with the initial conditions: $\hat{\omega} = \hat{a} = 0$ at $\tau = \tau_0 = k_z(t_0) / k_y$. Note that the precise value of the initial conditions is irrelevant as we are mainly interested in the long time limit where the memory of the initial conditions has been lost.
 
We are also interested in the transport of chemical species whose concentration is again split in a large-scale and a fluctuating part: $N = N_0 + n$. In the quasi-linear approximation, by assuming the molecular diffusivity of particle to be the same as the molecular viscosity \citep[for the case of non-unit Prandtl number, see][]{Kim06}, we obtain the following equation for the fluctuating density of particles:
\begin{equation}
\label{ParticleTransp}
\partial_\tau \hat{n} = i \frac{\hat{\omega}}{k_y (1+\tau^2)} \partial_z N_0 \; ,
\end{equation}
which is the same as the second equation in (\ref{System2}) up to a multiplicative constant. Thus the turbulent diffusivity of magnetic field is the same as the turbulent particle diffusivity.

\section{Hydrodynamical results}
\label{Hydrodynamics}
We first consider the case without a magnetic field ($\gamma = 0$) to elucidate the influence of Rossby waves on a sheared turbulence. Equation (\ref{System2}) with $\gamma=0$ can be readily integrated with the following result for the fluctuating vorticity:
\begin{equation}
\label{SolutionHydro}
\hat{\omega} = \int_{\tau_0}^\tau \frac{dt}{\A} e^{i\alpha [\arctan\tau - \arctan t]} \hat{f}_w(t) \; . 
\end{equation}
To calculate the turbulence intensity and the turbulent transport of angular momentum (\ie the eddy viscosity $\nu_T$ defined as $\langle v_y v_z \rangle = - \nu_T \partial_z U_0 = \nu_T \A$), we prescribe the statistical properties of the forcing $f_w$ to be spatially homogeneous and temporally short correlated with the correlation time $\tau_f$:
\begin{equation}
\label{Forcing}
\langle \tilde{f}_w({\bf k_1},t_1) \tilde{f}_w({\bf k_2},t_2) \rangle = \tau_f \, (2\pi)^2 \delta({\bf k_1}+{\bf k_2}) \, \delta(t_1-t_2) \, \phi({\bf k_2}) \; .
\end{equation}
Here, $\phi({\bf k_2})$ is the power spectrum of the forcing. We focus on the strong shear limit: $\xi = \nu k_y^2 / \A \ll 1$, which is valid on characteristic scales of fluctuation $L \gg L_s = 2 \pi (\nu / \A)^{1/2}$. The calculation of the turbulence intensity and turbulent viscosity is similar to that of \citet{Kim05} which we refer the readers to for technical details. The following solutions are obtained:
\EQA
\label{ResultHydro}
\langle \omega^2 \rangle &\sim& \frac{\tau_f}{(2 \pi)^2 \A} \int d^2 {\bf k} \; \phi({\bf k})\frac{1}{3} \Bigl(\frac{3}{2\xi}\Bigr)^{1/3} \Gamma(1/3) \; , \\ \nonumber
\nu_T &\sim& - \frac{\tau_f}{2 (2\pi)^2 \A^2} \int d^2 {\bf k} \; \frac{\phi({\bf k})}{k^2} \; . 
\ENA
The results in Eq. (\ref{ResultHydro}) are the same as those in the 2D turbulence with $\beta = 0$ \citep{Kim01}. That is, turbulence amplitude and transport are reduced only by shear stabilization,  and Rossby waves have no influence on the intensity of turbulence or the transport of angular momentum. In the kinematic regime ($\gamma=0$), we can also compute the magnetic diffusivity $\eta_T$, which enters the induction equation for the mean magnetic field and is defined as $\langle a v_z  \rangle = - \eta_T \partial_z A_0 = - \eta_T B_0$:
\begin{equation}
\label{TransportParticules}
\eta_T \sim \frac{2 \tau_f}{(2\pi)^2 \A^2} \int d^2 {\bf k} \; \frac{\phi({\bf k})}{\alpha^2 k_y^2} \sin^2\Bigr[\alpha \arctan \frac{1}{a}\Bigr] \sim \frac{2 \tau_f}{(2\pi)^2 \beta^2} \int d^2 {\bf k} \; \phi({\bf k}) \sin^2\Bigr[\alpha \arctan \frac{1}{a}\Bigr]\; .
\end{equation}
Here $\alpha = \beta / \A k_y$  and $a=k_z / k_y$. Thus, in the kinematic regime, the magnetic field diffuses more slowly compared to the 2D HD case (the sine cardinal function being always smaller than $1$). The last expression in Eq. (\ref{TransportParticules}) even shows that for large enough $\beta$, the magnetic diffusivity is reduced by a factor $\beta^{-2}$. Note that the same reduction applies to the transport of particle as $\eta_T = D_T$.

\section{MHD turbulence on a $\beta$ plane}
\label{MHD}
Combining the two equations of system (\ref{System2}), we can obtain the following equation for the magnetic potential:
\begin{equation}
\label{EquationMHD}
\partial_\tau\bigl[(1+\tau^2)\partial_\tau \hat{a}\bigr] - i \alpha \partial_\tau \hat{a} + \gamma^2 (1+\tau^2) \hat{a} = \frac{i \gamma}{\A k_y^2} \hat{f_w}
\end{equation}
The exact solution to Eq. (\ref{EquationMHD}) can be written in terms of Heun's confluent function; however, we here derive the physically more transparent results in the case of a weak and a strong magnetic field separately.  The boundary between the weak ($\vert \gamma \vert \gg 1$ where $\gamma = B_0 k_y / \A$) and the strong magnetic regime ($\vert \gamma \vert \ll 1$) is given by $L_B = 2 \pi  B_0 / \A$.

\subsection{Weak magnetic field}
For  $\vert \gamma \vert \ll 1$, we calculate the correction to the magnetic potential and vorticity up to second order in $\gamma$ as:
\begin{equation}
\hat{\omega}(\tau) = \hat{\omega}_0(\tau) + \gamma \hat{\omega}_1 (\tau) + \gamma^2 \hat{\omega}_2(\tau) + \dots \qquad \mathrm{and} \qquad
\hat{a}(\tau) = 0 + \gamma \hat{a}_1(\tau) + \gamma^2 \hat{a}_2(\tau) + \dots \; ,
\end{equation}
where $\hat{\omega}_0(\tau)$ is given by (\ref{SolutionHydro}). Solving order by order, we find that $\hat{\omega}$ is an even function of $\gamma$ whereas $\hat{a}$ is an odd function. The first correction to the hydrodynamical result is:
\EQA
\label{SolutionExp}
\hat{a}_1(\tau) &=& \int_{\tau_0}^\tau dt \; \frac{\hat{f}_w(t)}{k_y^2 \A \alpha} \Bigl(e^{i\alpha[\arctan\tau - \arctan t]}-1\Bigr) \; , \\ \nonumber
\hat{\omega}_2(\tau) &=& \frac{i}{\A \alpha} \int_{\tau_0}^\tau dt \; \hat{f}_w(t) e^{i\alpha\arctan\tau} \int_t^\tau dt' \; \Bigl(e^{-i\alpha\arctan t} - e^{-i\alpha\arctan t'}\Bigr)\; .
\ENA
Using Eqs. (\ref{SolutionExp}), we can easily calculate the second order (in $\gamma$) correction to the turbulence amplitude and turbulent stresses. Assuming that the forcing has the following symmetry property $\phi(k_y) = \phi(-k_y)$, we find that the first correction to the turbulence amplitude vanishes as it is odd in $k_y$. The turbulence amplitude is thus given by Eq. (\ref{ResultHydro}), showing that neither Alfv\'en nor Rossby waves change the turbulence amplitude. On the contrary, the Reynolds and Maxwell stresses are modified as follows:
\EQA
\nonumber
\langle v_y v_z \rangle &\sim& - \frac{\tau_f}{(2\pi)^2 \A} \int d^2 {\bf k} \; \phi({\bf k}) \Bigl\{\frac{1}{2 k^2} + \frac{2 \gamma^2}{k_y^2  \alpha^2} \sin^2\Bigl[\frac{\alpha}{2} \arctan \frac{1}{a} \Bigr] \Bigr\} \; , \\ \nonumber 
\langle b_y b_z \rangle &\sim& - \frac{4 \tau_f}{3 (2\pi)^2 \A} \int d^2 {\bf k} \; \frac{\gamma^2 \phi({\bf k})}{k_y^2 \alpha^2}  \sin^2\Bigl[\frac{\alpha}{2} \arctan \frac{1}{a} \Bigr] \Bigl(\frac{3}{2\xi}\Bigr)^{2/3} \Gamma(2/3) \; . 
\ENA
Here, $\alpha = \beta / \A k_y$, $a=k_z / k_y$ and $\xi = \nu k_y^2 / \A$. One can see that the correction to the magnetic stress is much larger than that to the Reynolds stress (recall that $\xi \ll 1$). Comparing the zeroth order solution for the Reynolds stress and second order for the Maxwell stress, we find that the former is dominant if $\gamma^2 \alpha^{-2} \xi^{-2/3} \ll 1$ (giving us the result of the 2D case). In the opposite case ($\gamma^2 \alpha^{-2} \xi^{-2/3} \gg 1$), the transport of angular momentum is mainly controlled by the magnetic part of the stress \ie by Alfv\'en wave turbulence. In particular, the turbulent viscosity $\nu_T = (\langle v_y v_z \rangle - \langle b_y b_z \rangle)/ \A$ is now positive, reminiscent of the direct cascade of energy in MHD turbulence \citep{Kim01}. These results enable us to calculate the characteristic scale at which the transition from the Rossby wave turbulence (with negative viscosity) to the Alfv\'en turbulence (positive viscosity) occurs, $L_R = 2 \pi (\A B_0^3 / \beta^3 \nu)^{1/4}$, which corresponds to the \lq\lq Rhines scale\rq\rq~(referring to the scale which separates, in ordinary geostrophic turbulence, eddy turbulence from wave turbulence). Our result shows that $L_R \sim R_e^{1/4} \sim R_m^{1/4}$ as we assume a unit Prandtl number. However, it is easy to show that for $\eta \gg \nu$ (as is relevant for the tachocline) or equivalently for $P_m \ll 1$, the first correction to the magnetic potential is the same as previously. Consequently, the magnetic stress remains the same, with $\nu$ being replaced by $\eta$. Thus, for low Prandtl number, the crossover scale is  $L_R = 2 \pi (\A B_0^3 / \beta^3 \eta)^{1/4} \sim R_m^{1/4}$. This scaling is the same as that of \citet{Diamond06}  obtained in a slightly different context (MHD turbulence on a $\beta$ plane without a shear flow).

In a similar way, we can calculate the turbulent diffusivity $\eta_T$. To calculate the first two leading order terms, we need to keep terms up to the third order in $\gamma$ for the magnetic potential (not shown here for brevity). Here again, only the leading order term survives when assuming $\phi(k_y) = \phi(-k_y)$ and thus the turbulent diffusivity remains the same as the one obtained in the hydrodynamical case [see Eq. (\ref{TransportParticules})].

\subsection{Strong magnetic field}
When $\vert \gamma \vert \gg 1$, we seek for a WKB solution of Eq. (\ref{EquationMHD}). Up to second order in $\epsilon = 1/\gamma$, a solution for the homogeneous part can be found as:
\begin{equation}
F(\tau) = A \frac{1}{\sqrt{1+\tau^2}} \exp\Bigl[i \frac{\alpha}{2} \arctan \tau + \epsilon^2 H(\tau) \Bigr]   \cos\Bigl[\frac{1}{\epsilon} G(\tau) + \phi \Bigr] \; .
\end{equation}
Here, $A$ and $\phi$ are two integration constants:
\EQA
G(\tau) &=& \tau + \frac{\epsilon^2}{4} \Bigl(\frac{\alpha^2}{4} - 1\Bigr) \Bigl(\arctan \tau + \frac{\tau}{1+\tau^2}\Bigr) \; , \\ \nonumber
H(\tau) &=& \Bigl(1 - \frac{\alpha^2}{4} \Bigr)\frac{1}{4(1+\tau^2)^2} \; .
\ENA 
Using the method of variation of constants, one obtains the solution for the magnetic potential and the vorticity:
\EQA
\label{SolutionWKB}
\hat{a}(\tau) &=& \int_{\tau_0}^\tau dt \; \frac{i \hat{f}_w(t)}{\A k_y^2 G'(t) \sqrt{1+t^2} \sqrt{1+\tau^2}} \exp\Bigl[\frac{i\alpha}{2} \Bigl(\arctan \tau - \arctan t\Bigr) \\ \nonumber
&& \qquad + \epsilon^2 \bigl(H(\tau)-H(t)\bigr)\Bigr] \sin\frac{1}{\epsilon}\bigl\{G(\tau)-G(t)\bigr\} \; , \\ \nonumber
\hat{\omega}(\tau) &=& \int_{\tau_0}^\tau dt \; \frac{\hat{f}_w(t) \sqrt{1+\tau^2}}{\A G'(t) \sqrt{1+t^2}} \exp\Bigl[\frac{i\alpha}{2} \Bigl(\arctan \tau - \arctan t\Bigr) + \epsilon^2 \bigl(H(\tau)-H(t)\bigr)\Bigr] \times \\ \nonumber
&& \qquad \Bigl[G'(\tau) \cos \frac{1}{\epsilon} \bigl\{G(\tau)-G(t)\bigr\} + \frac{i \alpha G'(t)^2 - 2 \tau}{2(1+\tau^2)G'(\tau)^2} \epsilon  \sin\frac{1}{\epsilon}\bigl\{G(\tau)-G(t)\bigr\} \Bigr] \; .
\ENA
We can then use Eqs. (\ref{SolutionWKB}) to calculate the Reynolds and Maxwell stresses and the turbulent viscosity. In the strong shear limit ($\xi \ll 1$), we obtain the following:
\begin{equation}
\label{Turbdiff}
\nu_T = \frac{\tau_f}{4 (2 \pi)^2 B_0^2} \int d^2{\bf k} \; \frac{\phi({\bf k}) k_y^2}{k^6} \Bigl(1-\frac{\alpha^2}{4}\Bigr) \; .
\end{equation}
Equation (\ref{Turbdiff}) shows that the turbulent viscosity is the same as in 2D MHD \citep{Kim01} except for the additional term proportional to $\alpha^2$ due to the transport reduction by Rossby waves. However, the strong magnetic field limit is only valid for $k_y \ll \A / B_0$, and thus the correction term is bounded from above: $\alpha \ll \beta B_0 / \A^2$. Using typical solar values, $B_0 \sim 1 \; \mbox{T}$, $\A \sim 6 \times 10^{-8} \; \mbox{s}^{-1}$, $\beta \sim 2.7 \times 10^{-15} \; \mbox{m}^{-1} \mbox{s}^{-1}$, we obtain $\alpha \ll 0.75$ and consequently the correction term (proportional to $\alpha^2$) is very small. 
 
The turbulent diffusivity can also be obtained in a similar way:
\EQA
\eta_T = \frac{\tau_f}{6 (2 \pi)^2 B_0^2} \int d^2{\bf k} \; \frac{\phi({\bf k})}{k^2 k_y^2} \Bigl(\frac{3}{2}\Bigr)^{1/3} \Gamma(1/3) {\Bigl(\frac{\nu k_y^2}{\A}\Bigr)^{2/3} } \; ,
\ENA
which is the same as the 2D case. Therefore, both the turbulent viscosity and diffusivity are reduced by a strong large-scale magnetic field while the effect of Rossby waves is negligible. 

\section{Discussion}
\label{Conclusion}
In order to elucidate the turbulent transport in the tachocline, we considered a 2D model of turbulence on a $\beta$ plane, in the presence of a large-scale latitudinal shear (with shearing rate $\A$) and a uniform toroidal magnetic field (intensity $B_0$). By using the quasi-linear approximation, we have computed the turbulent viscosity and diffusivity (of magnetic fields or particle) in the two limiting cases of weak and strong magnetic field. The intensity of turbulence is not altered much by the presence of waves whereas the turbulent transport can be severely affected due to the effect of these waves on the phase of the field. 

In the case of a weak magnetic field, we have shown that there is a crossover scale $L_R = 2 \pi (\A B_0^3 / \beta^3 \eta)^{1/4}$ from an Alfv\'en dominated (on small scales) to a Rossby dominated turbulence (on scale large enough). In the Rossby dominated regime, we found a negative turbulent viscosity, suppressed only by the shear: $\nu_T \propto - \A^{-2}$. In the Alv\'en dominated regime, the turbulent viscosity is positive with the following scaling: $\nu_T \propto B_0^2 (\beta \A)^{-2}$. We also have shown that the turbulent diffusivity is not very much affected by a weak magnetic field and that for all scales, the turbulent diffusivity  is suppressed only by Rossby waves: $\eta_T \propto \beta^{-2}$ (the same as in the HD case). In the case of a strong magnetic field, we found that the turbulent viscosity and diffusivity are positive and the same as in the 2D MHD case (except for an additional factor in the turbulent viscosity which is very small in the tachocline context) and consequently are suppressed only by magnetic fields: $\eta_T \propto \nu_T \propto B_0^{-2}$.

\begin{figure}[h]
\epsscale{.80}
\plotone{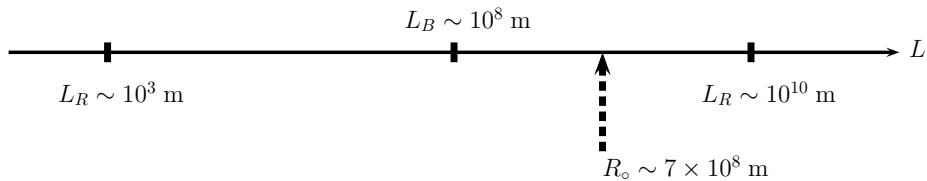}
\caption{\label{TachoScale} Sketch of the different relevant scales evidenced in this paper for tachoclinic values of the parameters. $L_R$ is the scale which separates the strong magnetic field regime from the weak one and $L_B$ (obtained in the weak field approximation) is the scale which separates a Alfv\'en dominated regime from a Rossby dominated one. We also included the smallest scale $L_S$ for which the large shear limit is valid.}
\end{figure}

These findings have interesting implications for the dynamics of the tachocline. Using typical solar values, $B_0 \sim 1 \; \mbox{T}$, $\A \sim 6 \times 10^{-8} \; \mbox{s}^{-1}$, $\beta \sim 2.7 \times 10^{-15} \; \mbox{m}^{-1} \mbox{s}^{-1}$ , $\nu \sim 10^{-2} \mbox{m}^2 \mbox{s}^{-1}$ and $\eta \sim 1 \mbox{m}^2 \mbox{s}^{-1}$, we can calculate the boundary between the weak and the strong magnetic regime, $L_B = 2 \pi  B_0 / \A \sim 10^8 \; \mbox{m}$ and the scale at which the transition from Alfv\'en to Rossby turbulence occurs, $L_R = 2 \pi (\A B_0^3 / \beta^3 \eta)^{1/4} \sim  10^{10} \; \mbox{m}$. Let us recall that this scale was calculated in the weak magnetic field case ($L > L_B$) which is consistent in the tachoclinic case as $L_R > L_B$. Figure \ref{TachoScale} shows the relative positions of these two scales compared to the radius of the Sun. We also show on this figure the minimal scale $L_s = 2 \pi (\nu / \A)^{1/2} \sim 10^3 \; \mbox{m}$ for which our strong-shear approximation holds. 
As relevant motion in the tachocline are probably on scales smaller than the solar radius, the dynamics of the tachocline is likely to be insensitive to the effect of Rossby waves. In the worst case, the dynamics would be that of the weak magnetic field with Alfv\'en dominated turbulence (meaning a positive turbulent viscosity). These results are mainly due to the enhancement of the back-reaction of fluctuating magnetic fields by shearing.

The reduction of the turbulent diffusion and transport induced by Alfv\'en waves may be problematic in the solar context, invalidating some of the mechanisms believed to be important for the tachocline dynamics \citep[see also][]{Diamond06}, e.g. the turbulent horizontal viscosity model of \citet{Spiegel92}. However, the important point in this particular scenario is that the horizontal viscosity can be much larger than the vertical one. We have shown \citep{2Shears} that it is the case for a turbulence sheared more strongly in the radial direction than in the latitudinal one (as is the case in the tachocline where the radial shear is at least one order of magnitude greater than the latitudinal one). If the effect of the magnetic field is the same in the two directions (reduction by a factor $B_0^2$), this result should not be altered by the inclusion of a magnetic field. However, a detail analysis of this situation requires the study of a three-dimensional model and is outside the scope of this paper.  

Finally, we note that in this paper, we assumed the toroidal magnetic field to be uniform on a local Cartesian ($\beta$) plane. However, on a sphere, there is a possibility of instability in presence of a band of toroidal magnetic field and latitudinal differential rotation \citep{Gilman97}. The turbulence arising from this instability can then be considered as a source of part of the forcing in our model.
 


\acknowledgements
We thank D. W. Hughes and S. M. Tobias for useful comments. This work was supported by U.K. PPARC Grant No. PP/B501512/1.

\end{document}